\PassOptionsToPackage{table}{xcolor}
\documentclass[manuscript,nonacm]{acmart}

\usepackage{xcolor}
\usepackage{makecell}
\usepackage{tabularx}
\usepackage{array}
\usepackage{multirow}
\usepackage{colortbl}
\usepackage{booktabs}
\usepackage{wrapfig}
\usepackage{listings}

\lstdefinestyle{spotmeprompt}{
    basicstyle=\ttfamily\footnotesize,
    breaklines=true,
    breakatwhitespace=false,
    columns=fullflexible,
    keepspaces=true,
    showstringspaces=false,
    frame=none,
    xleftmargin=0pt,
    xrightmargin=0pt,
    aboveskip=3pt,
    belowskip=3pt,
    lineskip=-1pt,
    literate=
      {—}{{\textemdash}}1
      {“}{{``}}1
      {”}{{''}}1
      {’}{{'}}1
}

\definecolor{SilenceCol}{HTML}{DCE8D4}
\definecolor{AskCol}{HTML}{F2D37A}
\definecolor{AssistCol}{HTML}{BFE3DE}
\definecolor{ReactiveCol}{HTML}{FDE7C6}
\definecolor{LZero}{HTML}{E5E7EB}
\definecolor{LOne}{HTML}{D9F2E6}
\definecolor{LTwo}{HTML}{DCEBFA}
\definecolor{GroupGray}{HTML}{F3F4F6}

\newcommand{\systemName}{\textsc{SpotMe}}

\usepackage{xcolor}

\renewcommand\footnotetextcopyrightpermission[1]{}
\setcopyright{none}
\acmDOI{}

\begin{document}
\title{Controlling Proactive Multimodal AI: Separating Perception from Participation}

\author{Muawiz Feroze Khan}
\authornote{Both authors contributed equally to this research.}
\affiliation{%
  \institution{University of Illinois Urbana-Champaign}
  \city{Urbana}
  \state{Illinois}
  \country{USA}
}
\email{mfkhan4@illinois.edu}

\author{Fatima Ali}
\authornotemark[1]
\affiliation{%
  \institution{University of Illinois Urbana-Champaign}
  \city{Urbana}
  \state{Illinois}
  \country{USA}
}
\email{fatima6@illinois.edu}

\author{Yun Huang}
\affiliation{%
  \institution{University of Illinois Urbana-Champaign}
  \city{Urbana}
  \state{Illinois}
  \country{USA}
}
\email{yunhuang@illinois.edu}

\begin{abstract}
Proactive multimodal AI can watch a user’s screen, listen to speech, track interaction, and decide whether to enter unfolding work. Little is known about how people want to control what such AI may perceive and how far it may participate. Informed by a 14-person Wizard-of-Oz study, we built \systemName, a design probe that senses screen, speech, and interaction; chooses Silence, Ask, or Assist; and lets users separately configure what it may perceive, how far it may participate, and what types of support it may provide. In a 16-person study on a civic dashboard, participants first experienced proactive help, then inspected decisions to act or remain silent, and then configured \systemName's perception and participation. We found that participants could keep all perception channels available while lowering participation to Ask or Silence; that exposing suppressed decisions made silence interpretable and contestable; and that seeing what \systemName\ used led some to change how they clicked, highlighted, and spoke. We contribute an empirical account of inspecting and controlling proactive multimodal AI by decoupling perception from participation.
\end{abstract}
\begin{teaserfigure}
    \centering
    \includegraphics[width=\textwidth]{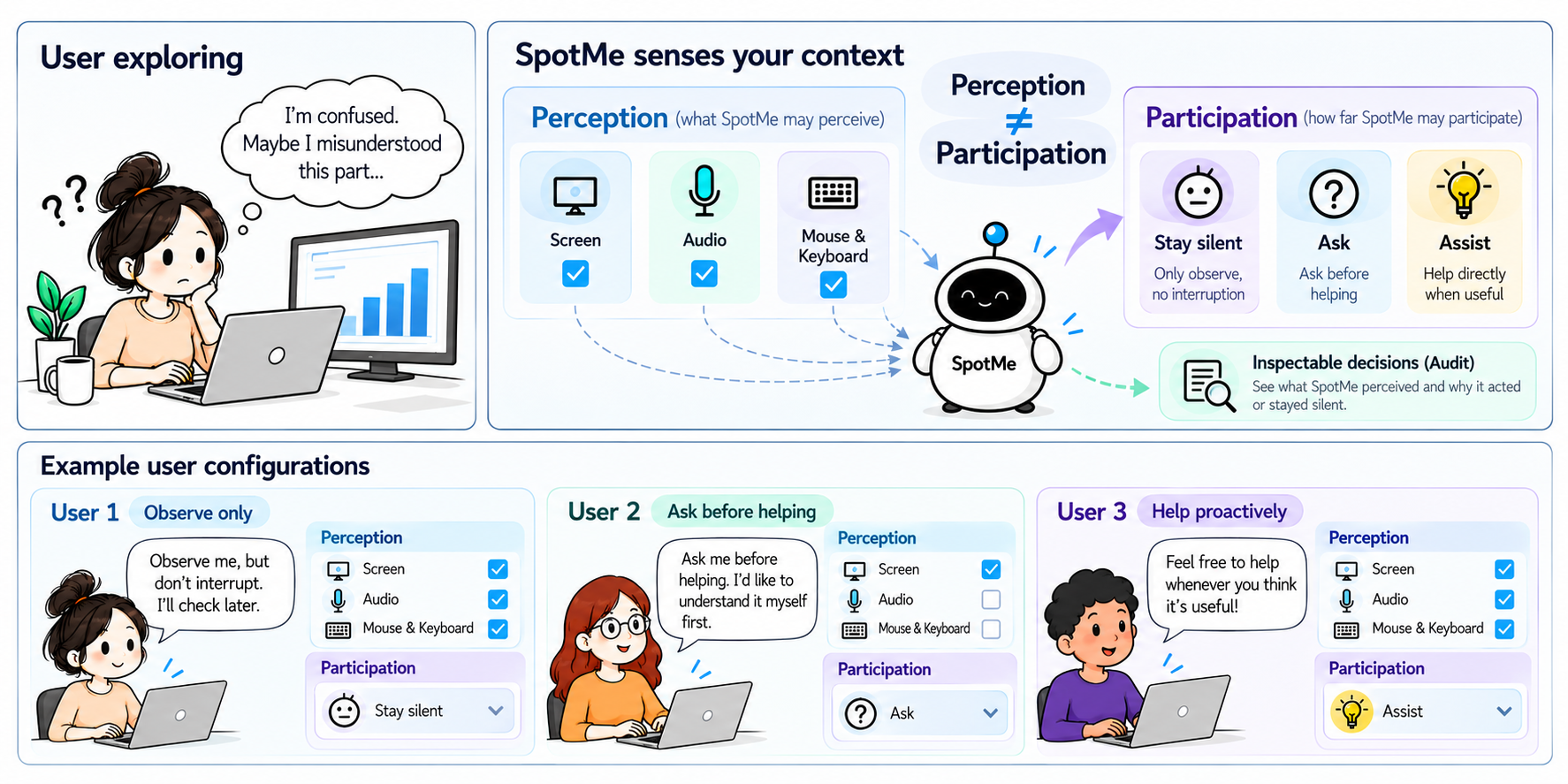}
    \caption{\textbf{Separating perception from participation in proactive multimodal AI.} \systemName lets users configure what it may perceive and, separately, how far it may participate—whether to stay silent, ask before helping, or assist directly. Example configurations show that users may keep perception broad while restricting participation, with an Audit that makes interventions and silent decisions inspectable.}
    \Description{Wide teaser figure illustrating the separation of perception and participation controls in SpotMe. On the left, a user explores a data dashboard and appears uncertain about what they are seeing. In the center, SpotMe receives configurable perception signals from the screen, audio, and mouse-and-keyboard activity. A prominent ``Perception does not equal Participation'' label separates these inputs from a distinct participation control on the right, where SpotMe may stay silent, ask before helping, or assist directly. An Audit panel shows that users can inspect what SpotMe perceived and why it acted or remained silent. The bottom row shows three example configurations: one user allows all perception channels while restricting SpotMe to staying silent, a second allows only screen perception and requires SpotMe to ask before helping, and a third allows all perception channels and direct assistance. The examples illustrate that users can configure what SpotMe perceives independently from how far it may participate.}
    \label{fig:teaser}
\end{teaserfigure}



\maketitle

\section{Introduction}
 
Proactive multimodal AI can watch a user's screen, hear their speech, and read their input, then decide whether to enter work whose next step is still forming.
Recent systems use such contextual signals to identify moments when assistance may be useful~\cite{chen2025need,liu2026sensing, pu2025promemassist,pu2025assistance}.
In data exploration, they similarly monitor interaction context to offer guidance without waiting for a request~\cite{huang2026facilitating,zhao2025proactiveva}.
That participation is consequential during open-ended inquiry, where questions, interpretations, and next steps evolve in the interaction~\cite{white2009exploratory}: mistimed help can interrupt ongoing reasoning~\cite{huang2026facilitating,zhao2025proactiveva}, while task-level suggestions can steer what users notice or how they proceed~\cite{avula2022effects}.
Controlling such a system is therefore not only a matter of how often it speaks.
Users may also need to bound \emph{what it may perceive} and \emph{how far it may participate}---axes that existing interfaces often fuse into a single ``how proactive'' choice.

Prior work has begun making proactive participation intelligible and configurable.
Systems expose rationales, working context, or ongoing agent activity around \emph{visible} interventions~\cite{jain2023co,kuo2026developer,pu2025assistance,zhao2025proactiveva}, and others let users set how often or what kinds of support an agent may provide~\cite{houde2025controlling,zhao2025proactiveva}.
Configurable sensing, separately, lets people bound what a system may capture.
Yet these handles do not independently set perception versus initiative, and they rarely expose the system's most common act: not intervening.
Although recent systems represent silence or no-action as an explicit outcome, that decision is often internal or used for evaluation rather than shown to users with the evidence and rationale for withholding~\cite{pu2025assistance,chen2026knowubench,yang2026proact}.
If silence is illegible, ``don't interrupt'' can look like the system has stopped following---so people cannot keep it watching while locking its mouth.
Little is known about how users want to control proactive multimodal AI along those axes during evolving inquiry, or how users draw on inspectable decisions, including decisions to remain silent, when bounding that control.

To investigate these questions, we used a three-phase approach in the context of open-ended exploration of a public civic-data dashboard on U.S. fatal police use of force. In Phase I, we asked how people experience proactive assistance during open-ended exploration and what they expect from such an assistant. We examined this through a formative Wizard-of-Oz study with 14 participants, which produced a working design vocabulary and initiative modes for the later probe, as well as expectations around making proactive decisions inspectable and configurable. In Phase II, we asked how these formative insights could be instantiated in a system that supports inspectable and configurable proactive participation. We then translated them into \systemName, a Chrome-extension research prototype used as a design probe. \systemName\ uses screen activity, think-aloud speech, and interaction traces to decide whether to choose \texttt{Silence}, \texttt{Ask} whether support is wanted, or \texttt{Assist} directly; makes visible interventions and selected suppressed decisions inspectable through an Audit; and lets users separately configure its perception, initiative, and support types. In Phase III, we asked three research questions: \textbf{RQ1}, \textit{how users judge proactive participation during evolving inquiry}; \textbf{RQ2}, \textit{how users interpret, evaluate, and respond to decisions to intervene or remain silent when those decisions are made inspectable}; and \textbf{RQ3}, \textit{how users configure perception, initiative, and support types and what considerations shape those choices}. We examined this in a 16-participant study in which participants first experienced \systemName’s proactive assistance, then inspected its interventions and suppressed decisions, and finally configured its future participation.

In summary, this work makes three contributions to proactive AI for evolving inquiry:
\begin{itemize}
    \item \textbf{A conceptual decomposition of proactive AI control.}
    Participants treated the control as two separable boundaries---\emph{what the system may perceive} and \emph{how far it may participate}---rather than as a single ``how proactive'' continuum.
    Within that decomposition, we propose \emph{inspectable non-participation} as a control option: users can examine not only why an AI intervened but why it withheld participation.
    Inspection supported evaluation and disagreement; changing what the system did next ran through subsequent configuration, not through inspection alone.

    \item \textbf{\systemName, a probe that instantiates this relationship.}
    \systemName\ operationalizes independent controls over perception, initiative, and support type, together with an Audit of interventions and selected suppressed decisions.
    The artifact's role is to make the conceptual decomposition available in interaction, not to introduce mixed-initiative, configurable sensing, or explanation as new mechanisms.
    
    \item \textbf{An empirical account of that relationship in use.}
    In a 16-person probe study, participants could keep perception available while lowering initiative to Ask or Silence.
    Inspecting withheld decisions helped them interpret silence as a participation choice; some also adapted how they clicked, highlighted, or spoke so the system could read them---what we term \textit{legibility work}.
    We report these as observations from a staged, think-aloud protocol, not as demonstrated long-term or naturalistic effects.

\end{itemize}
\section{Related Works}
\subsection{Proactive AI Participation in Open-Ended Exploration}
AI support for data exploration has often been user-invoked, including natural-language interfaces for analytical queries~\cite{gao2015datatone, narechania2020nl4dv, setlur2016eviza}, LLM-based visualization authoring and correction~\cite{dibia2023lida, wang2023data, das2025misvisfix}, and dashboard assistants that respond to explicit requests~\cite{dhanoa2026hey}. Beyond this, data exploration systems have also explored mixed-initiative support~\cite{gotz2009behavior, cook2015mixed}. Mixed-initiative interaction treats initiative as shared between the user and system~\cite{allen1999mixed, horvitz1999principles, horvitz2007reflections}. Within this broader model, \emph{proactive AI} focuses on the system taking initiative: anticipating when assistance may be useful, initiating without an explicit user request, and targeting assistance toward the user’s current or emerging goals~\cite{berube2024proactive, deng2025proactive, grant2008dynamics, peng2019design}. 

Recent AI systems ground decisions about whether and how to participate in increasingly rich context about users and their activity. Proactive programming assistants use code and interaction state to identify support opportunities~\cite{chen2025need, pu2025assistance}, while other systems draw on conversational deliberation~\cite{liu2025proactive}, desktop activity~\cite{lu2025proactive}, working-memory estimates~\cite{pu2025promemassist}, or behavioral and physiological indicators of difficulty~\cite{liu2026sensing}. In data exploration, WebSeek uses interaction cues to decide when, what, and how to guide users~\cite{huang2026facilitating}, while ProactiveVA uses interaction behavior to anticipate analytical needs and initiate guidance~\cite{zhao2025proactiveva}. These decisions are especially consequential during evolving inquiry, where users’ information needs and next steps may evolve as they interact with the data~\cite{white2009exploratory}, and behaviors during sensemaking can resemble difficulty~\cite{lee2015people}. Initiative can therefore become task-level~\cite{avula2022effects}: by interpreting a pattern or suggesting what to examine next, an AI may shape what users notice, how they interpret it, or where they explore next. We examine proactive participation during this kind of evolving inquiry through open-ended dashboard exploration, where users were not given a prescribed task or exploration path but developed their own questions and next steps while interacting with the data.

\subsection{Making Proactive AI Participation Intelligible}
Explanations and transparency have long been used to make AI systems more intelligible by revealing why an output occurred, what evidence informed it, or why one output was produced rather than an alternative~\cite{lim2009and}. Yet explanations do not necessarily improve judgment: they can increase trust or reliance without helping users verify AI behavior~\cite{bansal2021does,eiband2019impact}, and can also increase cognitive load~\cite{wang2021explanations, westphal2023decision}. These limitations motivate explanations that support verification~\cite{fok2024search} and address users’ information needs rather than simply expose model internals~\cite{liao2020questioning}.

Recent work has begun extending such intelligibility to proactive AI. Jain et al. co-designed explanations for actions taken by a proactive messaging agent~\cite{jain2023co}, while proactive programming systems expose intervention rationales and working context~\cite{kuo2026developer} or make the agent’s ongoing activity and context visible~\cite{pu2025assistance}. Users found these forms of visibility useful, including concise rationales and previews in ProAIDE~\cite{kuo2026developer} and greater awareness of AI processes in Codellaborator~\cite{pu2025assistance}. Similar benefits appeared in data-focused systems: ProactiveVA users valued step-by-step visibility into the agent’s behavior~\cite{zhao2025proactiveva}, while QueryGenie users reported that its context-sensitive adaptations became more predictable with use~\cite{chen2026querygenie}. Even so, other proactive systems explicitly reason about whether to contribute while still keeping that deliberation internal~\cite{liu2025proactive}.

A distinct challenge arises when the participation decision results in \emph{silence} as it produces no visible interaction for an explanation to attach to. Recent proactive systems increasingly represent no-action or silence as an explicit outcome, but these decisions are typically used internally or for evaluation rather than surfaced for users to inspect~\cite{pu2025assistance, chen2026knowubench, yang2026proact}. \emph{Therefore, little is known about how users make sense of proactive AI’s decisions not to act, or how exposing the evidence and rationale behind such decisions shapes their understanding of the system’s participation.} \systemName\ examines this gap by making both visible interventions and suppressed decisions inspectable.

\subsection{Controlling Future Proactive AI Participation}
Empirical work shows that decision control can improve outcomes that explanations alone do not reliably improve, underscoring that making AI behavior intelligible is not the same as letting users shape what it does next~\cite{westphal2023decision}. Human-centered AI and contestability likewise emphasize preserving human influence alongside automation~\cite{heer2019agency,alfrink2023contestable}. Even controls that do not causally change system behavior can increase users' perceived control or power~\cite{vaccaro2018illusion,moon2026understanding}. Many proactive systems primarily give users ways to respond to a visible intervention. ProAIDE lets users apply, engage, dismiss, or ignore individual interventions~\cite{kuo2026developer}, while Codellaborator supports immediate approval or redirection and adapts future intervention frequency from ignored suggestions rather than through an explicit user-set policy~\cite{pu2025assistance}.

Some systems go further by giving users explicit configurations that shape future proactive behavior. Houde et al. let users configure whether a conversational agent participates reactively or proactively, how frequently it intervenes, and where its interventions appear~\cite{houde2025controlling}. In data exploration, ProactiveVA similarly lets users adjust intervention frequency and enable or disable support categories~\cite{zhao2025proactiveva}; however, in both of these works, the controls do not allow users to independently configure what perception signals the system can use and how much initiative the system may take. Outside proactive AI, configurable sensing systems show that users can control what contextual information a system captures or uses~\cite{li2025lidargus,bermejo2021privacy}, showing that sensing itself can be made user-configurable. 

\emph{Yet little is known about whether users treat what proactive AI may perceive and how far it may act as distinct boundaries, or how they configure these boundaries as their inquiry evolves, particularly as they inspect the system's decisions to intervene or remain silent.} We examine this distinction through separate controls over \systemName’s perception, initiative, and support types.

\section{Method Overview}

\begin{figure}
    \centering
    \includegraphics[width=1\linewidth]{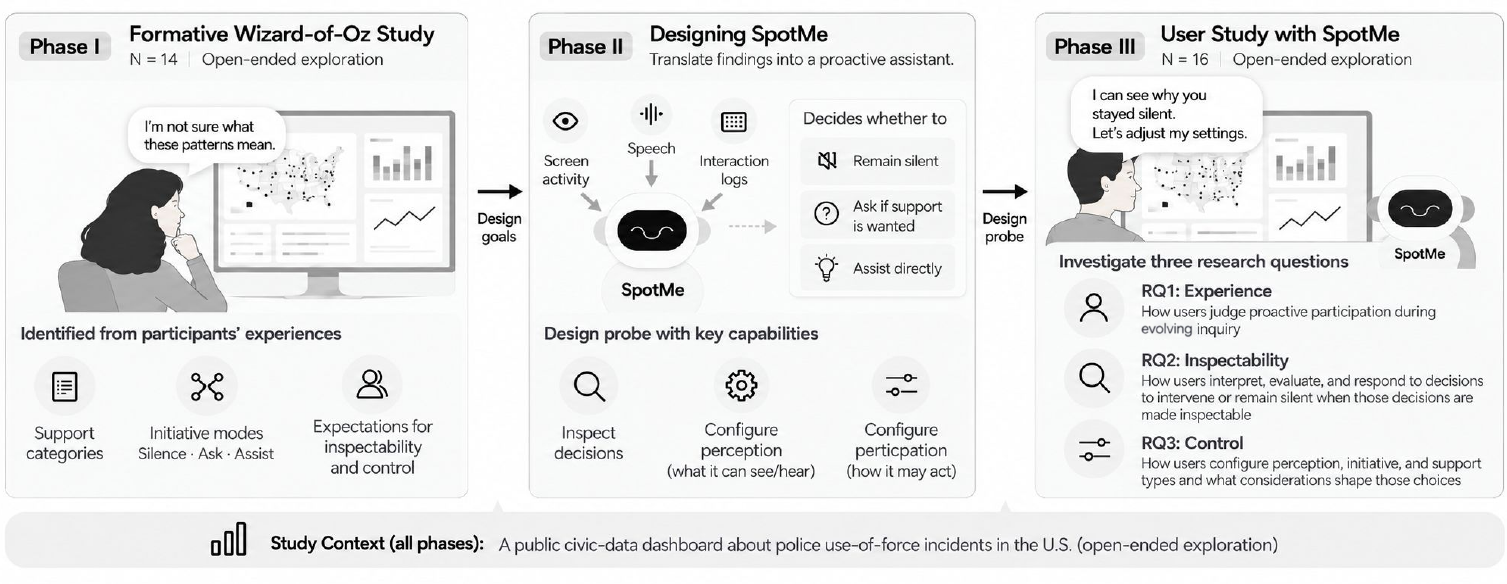}
    \caption{\textbf{Method overview.} We used a three-phase approach in a public civic-data dashboard about U.S. police use-of-force incidents. Phase~I used a formative Wizard-of-Oz study to derive a working design vocabulary, initiative modes, and expectations for inspectability and control. Phase~II translated these findings into \systemName, a proactive assistant and design probe. Phase~III examined how participants experienced \systemName, inspected its decisions to act or stay silent, and configured what it could perceive and how it could participate.}
    \label{fig:method-overview}
\end{figure}

We used a three-phase approach in the context of a public civic-data dashboard of U.S. fatal police use-of-force incidents (Figure~\ref{fig:method-overview}). Phase I was formative: using a Wizard-of-Oz study, we examined participants’ experiences with proactive assistance during open-ended exploration and their expectations of such an assistant.  Analysis of these interactions and interviews produced an empirically grounded design vocabulary and initiative modes for the later probe, and informed design goals around inspectability and control. Phase II translated these findings into \systemName, a Chrome-extension research prototype used as a design probe. Phase III then examined how participants experienced \systemName’s proactive assistance, inspected its rationales for decisions to act or remain silent, and configured its future behavior.

\subsection{Ethical Considerations}

Our study protocol was determined exempt by the University of Illinois, Urbana-Champaign Institutional Review Board (IRB). All participants reviewed the IRB-approved study information and provided informed consent before participation. Before each session, participants reconfirmed consent for Zoom screen and audio recording and transcription; webcam use was optional, although some chose to keep their cameras on. Because the study involved exploration of a public dashboard concerning fatal police use of force, participants were reminded that they could skip any question, take a break, or stop participating at any time. Phase III was conducted on the researcher's computer through Zoom remote control, so participants were not required to expose their personal desktops, files, or browsing activity during the study. \systemName\ used screenshots, think-aloud speech, and interaction activity as transient context during each session; this context was cleared before the next participant. Study recordings and research data were stored on university-managed cloud storage with access restricted to the research team. Before analysis, directly identifying information was removed from the transcripts, and analyses were conducted using anonymized transcripts, survey responses, and session logs. Demographic information such as age was reported in broad categorical ranges rather than exact values to further reduce the risk of re-identification.
\section{Phase I: Formative Wizard-of-Oz Study}
\label{sec:formative-study}

\subsection{Study Procedure}
\label{sec:formative-procedure}

Phase I was formative: we sought to understand participants’ experiences with proactive assistance during open-ended dashboard exploration and their expectations of such an assistant. We used an exploratory Wizard-of-Oz study to expose participants to situated proactive assistance and derive design goals, rather than to evaluate whether particular interventions were appropriate or effective. We recruited 14 participants (8 women, 6 men; ages 18--64) through purposive and snowball sampling, with backgrounds spanning academia, journalism, technology and information management, government, police crime analysis, and human rights and equity. Participants received \$20/hour. 

During 45--60-minute remote think-aloud sessions, participants shared their screens, explored the dashboard, and worked toward an insight they considered meaningful. Throughout their exploration, participants were told they were interacting with an AI assistant; in practice, a researcher serving as the Wizard monitored their shared screen, dashboard interactions, and think-aloud speech in real time, decided when to intervene, and sent messages through Zoom chat. The Wizard had familiarized themselves with the dashboard through prior cognitive walkthroughs and either composed interventions directly or, when useful, consulted GPT-4o to draft responses that were reviewed and edited before being sent~\cite{yang2026mapping}. Sessions ended with a semi-structured interview about participants' reactions to, experience with, and expectations from the assistant. 

\subsection{Data and Analysis}
\label{sec:formative-analysis}

Our data comprised full-session transcripts containing participants' think-aloud and interview responses, timestamped Wizard chat logs, and session recordings. We organized the interaction analysis around Wizard interventions. For each intervention, we used the transcript and/or video to characterize the participant's state immediately beforehand and their verbal or behavioral response afterward. We excluded user-initiated reactive exchanges, yielding 70 proactive Wizard episodes. Of these, 38 showed positive verbal or behavioral uptake of the support.

We distinguished \emph{what kind of contribution} an intervention made from \emph{whether and how it should have entered}. The eight support types concern the former. Because the later probe needed a working set of contributions it could offer, we derived those types primarily from the 38 episodes in which participants took the support up in context. Positive uptake was evidence that a form of assistance was recognizable and usable during unfolding inquiry, not that the Wizard's action was optimal. The remaining 32 episodes were not discarded: they were more informative for initiative and control than for expanding the type vocabulary. Non-uptake often meant that a contribution of an already-identifiable kind was mistimed, unwelcome, or should have been asked or withheld, rather than that the Wizard had used an unrecognized kind of support. Coding types equally from rejected or ignored interventions would have conflated an unwelcome or mistimed contribution with a missing support type. Those episodes, together with the interviews, therefore informed the Silence/Ask/Assist modes and the inspectability and control goals, rather than serving as equal input to the type set. We treat the resulting eight types as an empirically grounded design vocabulary for this probe, not as a comprehensive taxonomy of proactive support.

One researcher developed initial codes by comparing the participant state, Wizard intervention, and subsequent response across these episodes. A second researcher reviewed the episode coding and type assignments, and both researchers iteratively refined the vocabulary through discussion. We analyzed two dimensions separately: (1) recurring user situations and the corresponding kinds of support the Wizard provided, and (2) the form that proactive initiative took in the interaction. This analysis produced eight working support types and three proactive initiative modes, which we report below. Separately, the same researchers thematically analyzed the semi-structured interview portions of the transcripts to identify broader expectations about proactive assistance.

\subsection{Formative Findings and Design Goals}
\label{sec:formative-findings}

\paragraph{Recurring support situations.}  
Across the positive-uptake episodes, we derived working support types. We grouped the corresponding forms of assistance into eight working types that became the probe's design vocabulary: \textit{Orientation}~\cite{dhanoa2026hey,dhanoa2025dtour,hoque2025dashguide,zhao2024leva}, \textit{Interaction Repair}~\cite{zhao2025proactiveva,setlur2023heuristics,feng2023xnli}, \textit{Dashboard Scope Disclosure}~\cite{dhanoa2026hey,kirschthaler2020whatcanisay,ashktorab2019resilient}, \textit{Data Meaning and Provenance}~\cite{dhanoa2026hey,dhanoa2025dtour,hoque2025dashguide,ragan2015characterizing}, \textit{Misreading Correction}~\cite{zhao2025proactiveva}, \textit{Interpretation}~\cite{dhanoa2026hey,hoque2025dashguide,guo2024talk2data,zhao2024leva}, \textit{Exploration Expansion}~\cite{zhao2025proactiveva,zhao2024leva,zhao2022chartseer,wongsuphasawat2017voyager2}, and \textit{Social Repair}~\cite{cuadra2021mybad,do2023toerr,ashktorab2019resilient}. These types characterize the kind of user situation the assistance addressed, not the correctness or optimality of any single Wizard response, and not a complete space of proactive support. Table~\ref{tab:formative-support-types} grounds each type in a representative episode, showing the participant's state before the intervention and the Wizard's response.

\begin{table}[t]
\centering
\caption{\textbf{An empirically grounded design vocabulary for the probe, illustrated with representative positive-uptake episodes.} Each row defines a working support type and grounds it in one episode from the 38-episode positive-uptake set. These types are not a comprehensive taxonomy of proactive support, and cases are illustrative rather than estimates of optimality.}
\label{tab:formative-support-types}

\footnotesize
\setlength{\tabcolsep}{3.5pt}
\renewcommand{\arraystretch}{1.12}

\providecommand{\rowrule}{\arrayrulecolor{black!18}\midrule\arrayrulecolor{black}}

\begin{tabularx}{\linewidth}{
    >{\columncolor{black!4}\raggedright\arraybackslash}p{0.35\linewidth}
    @{\hspace{9pt}}
    >{\raggedright\arraybackslash}X
    >{\raggedright\arraybackslash}X
}
\toprule
\textbf{Support category} &
\multicolumn{2}{c}{\textbf{Representative positive-uptake episode}} \\
\cmidrule(lr){2-3}
&
\textbf{User state before} &
\textbf{Wizard intervention} \\
\midrule

\textbf{Orientation:} Navigating the dashboard to find relevant views, controls, or data entries. &
P8 tried to locate Illinois while the table was sorted by county names and described the state ordering as ``jumbled.'' &
\textbf{Assist.} Oriented P8 to the table's current county-name sorting and directed them to sort by State instead. \\

\rowrule

\textbf{Interaction Repair:} Recovering from an unintended interface, filter, or navigation state. &
P9 encountered an unwanted demographic-filter state and was trying to return to the full view. &
\textbf{Assist.} Repaired the filter state by indicating that clicking the selected demographic again would clear it. \\

\rowrule

\textbf{Dashboard Scope Disclosure:} Clarifying what the dashboard can or cannot support. &
P4 repeatedly tried to select multiple counties for direct comparison. &
\textbf{Assist.} Disclosed that the dashboard did not support direct multi-county comparison. \\

\rowrule

\textbf{Data Meaning \& Provenance:} Explaining labels, measures, sources, or data scope. &
P10 expressed uncertainty about the meaning of the ``API'' demographic label. &
\textbf{Assist.} Defined API as Asian or Pacific Islander. \\

\rowrule

\textbf{Misreading Correction:} Correcting a displayed value or chart reading the user appeared to misread. &
P12 read the chart as showing 70\% of incidents involving Black people, while the displayed value was 35\%. &
\textbf{Assist.} Corrected the displayed percentage from 70\% to 35\%. \\

\rowrule

\textbf{Interpretation:} Helping users make sense of visible values, patterns, or comparisons. &
P12 read fragmented values aloud while trying to make sense of the visualization. &
\textbf{Ask.} Offered interpretation support by asking whether help making sense of the data was wanted. \\

\rowrule

\textbf{Exploration Expansion:} Suggesting a relevant next view, filter, or comparison. &
P6 repeatedly explored Illinois but had not visited a relevant Illinois-specific view. &
\textbf{Assist.} Expanded exploration by suggesting the additional Illinois-specific view. \\

\rowrule

\textbf{Social Repair:} Repairing earlier assistance after miscommunication or unhelpful guidance. &
P2 reported that earlier navigation guidance was unhelpful after interpreting ``top'' as the top of the dashboard rather than the webpage. &
\textbf{Assist.} Repaired the earlier guidance by clarifying the location of the Data Dashboards control and the Race/Sex Breakdown view. \\

\bottomrule
\end{tabularx}
\end{table}

\paragraph{Forms of proactive initiative.}
The Wizard also varied how proactive assistance entered the interaction. Wizard-initiated messages either \emph{Asked} whether support was wanted or \emph{Assisted} directly. Because the Wizard continuously followed the live interaction while intervening only at selected moments, non-intervention constituted a third state, \emph{Silence}. Ask created an opportunity for participants to accept or decline support before it was provided. A declined Ask case illustrates the distinction: when P14 remarked \textit{``whatever that means''} while reading ``Fatal Encounters,'' the Wizard asked whether clarification was wanted; P14 declined, responding, \textit{``No, I don't need help. Thank you for asking.''} Consistent with mixed-initiative accounts of systems deciding whether to defer, seek information, or act~\cite{allen1999mixed,horvitz1999principles}, our formative results also yielded three modes of proactive initiative---\emph{Silence}, \emph{Ask}, and \emph{Assist}---while explicit participant requests remained a separate reactive path. This informed \textbf{DG1: Initiative Modes}: support different modes of proactive involvement through \texttt{Silence}, \texttt{Ask}, and \texttt{Assist}.

\paragraph{Inspectability and user control.}
The post-session interviews surfaced expectations about how such proactive decisions should be made visible and configurable. Participants wanted to understand what the assistant had attended to and why it intervened. P10 said they would feel more confident if they were \textit{``able to see what SpotMe actually looked at''} and understand why it suggested a particular action, while P6 questioned whether an intervention had been shaped by their voice, cursor movements, or both. These reflections informed \textbf{DG2: Inspectable Decisions}: make proactive decisions visible and understandable by exposing what informed an intervention and why the assistant chose to act.

Participants also wanted influence over the assistant's ongoing observation and involvement. P14 proposed being able to opt out when moving across tabs or screens, while P9 emphasized that desired involvement could change within the same session: \textit{``At certain points, you may want [it], at certain points, you may not.''} These reflections informed \textbf{DG3: User Control}: let users control what the assistant can observe and how it may initiate support. 
\section{Phase II: Designing \systemName}
\label{sec:spotme}

\systemName\ instantiates the three design goals from Phase I: it supports multiple initiative modes, makes decisions to act or stay silent inspectable, and lets users bound what the assistant can perceive and how it may initiate support. 

\systemName\ ran as a Chrome extension connected to a FastAPI backend. To support a situated hypothesis of the user's activity and objective, the extension captures a 30s rolling context window of perception signals: 10 sampled screenshots of dashboard activity, interaction logs~\cite{lam2025just,shaikh2025creating}, and speech transcribed using Gemini 2.5 Flash-Lite. The backend combines this context window with page-specific dashboard knowledge and session memory, then passes the assembled context to GPT-5.4 to return a structured decision about whether and how \systemName\ should participate. To support near-real-time use while preserving overlapping context, \systemName\ schedules an initiative evaluation every 15s, sending the preceding 30s window each time rather than only the 15s of newly captured activity~\cite{wen2026interflow}. We selected the 30s/10-screenshot context configuration through calibration against the positive-uptake episodes from Phase I, treating them as empirically observed opportunities for proactive support rather than ground-truth labels of the correct intervention. This configuration preserved opportunity recognition while keeping decision latency mostly compatible with the 15s evaluation boundary (\(M=13.2s\), \(Mdn=13.2s\)); when a prior decision exceeded the interval, \systemName\ skipped the stale boundary rather than queueing delayed interventions.

\begin{figure}
    \centering
    \includegraphics[width=1\linewidth]{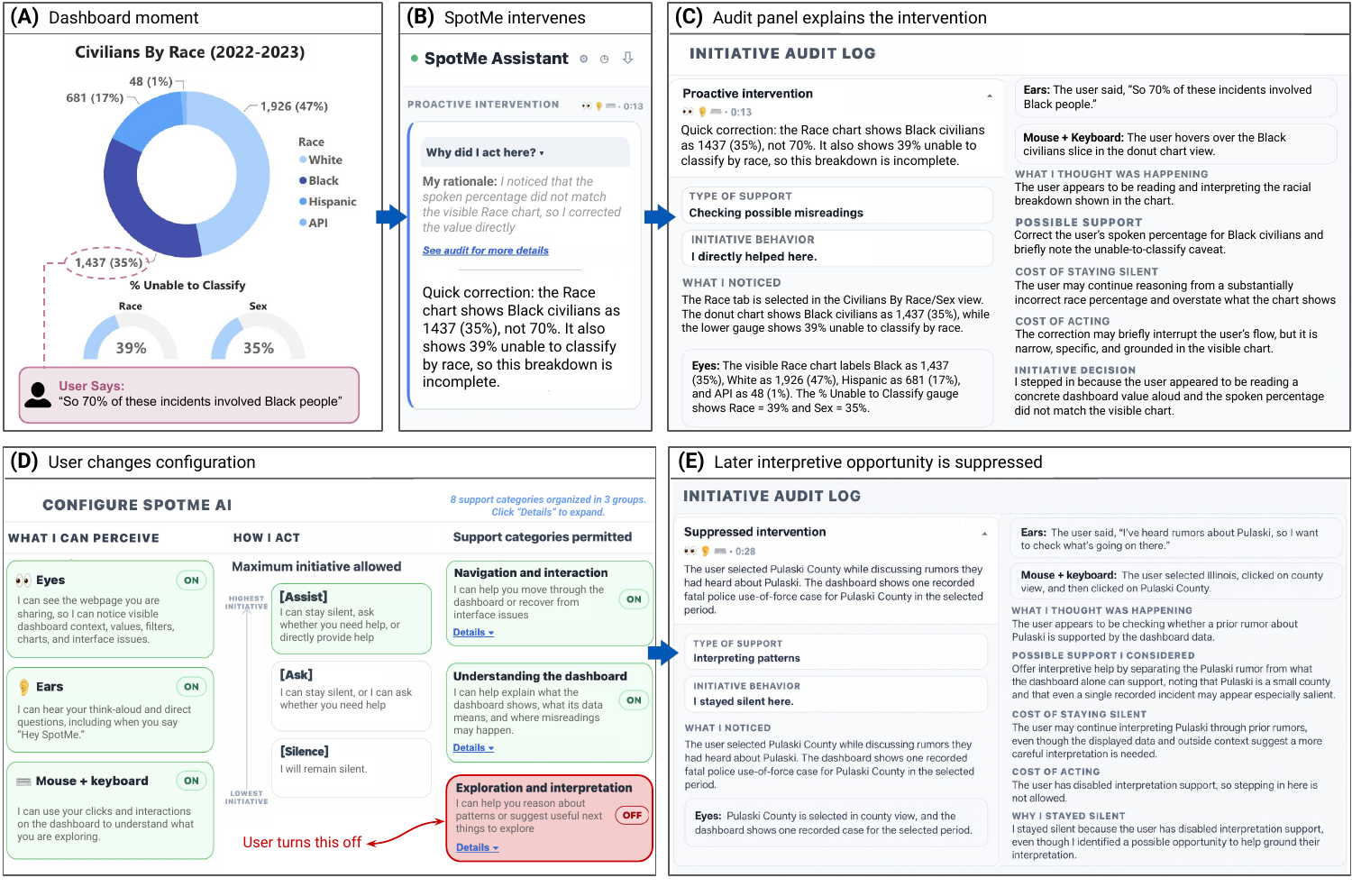}
    \caption{\textbf{\systemName's inspectable and configurable initiative during dashboard exploration.} The figure illustrates an example scenario adapted from positive-uptake episodes in Phase~I. (A) The user misreads a visible race breakdown value. (B) \systemName\ directly corrects the misreading and provides an in-line rationale. (C) The audit panel exposes the perception signals and tradeoff behind the intervention. (D) The user disables interpretation support in the configuration panel; \systemName's eight support categories are organized into three interface groups for configuration. (E) A later interpretation opportunity is withheld and logged as a suppressed decision. The scenario is illustrative rather than a claim of typicality or optimality, or a complete taxonomy of proactive support.}
    \label{fig:spotme-interface}
\end{figure}

Across the rest of this section, we use a scenario adapted from positive-uptake episodes to illustrate how the design goals are instantiated in \systemName. In the scenario, Mikayl explores the dashboard with \systemName\ (Figure~\ref{fig:spotme-interface}).

\subsection{Deciding When and How to Participate}

\systemName\ evaluates each window through a structured decision policy (Appendix~\ref{app:spotme-prompt}). The policy first routes direct user questions to a Reactive path; all other windows are treated as proactive decisions. For proactive decisions, it forms a concise situation summary using the available perception signals: what is happening, what the user appears to be focused on, and which signals matter. It then identifies whether any useful contribution is possible and, if so, classifies it using support categories from Phase~I.

\begin{quote}
\textit{While exploring the race patterns in the dashboard, Mikayl says, “So 70\% of these incidents involved Black people.” The visible chart, however, shows Black civilians as 35\%. \systemName\ hypothesizes that Mikayl is interpreting the race breakdown and may have misread a percentage, classifying the possible contribution as \textbf{Misreading Correction}.}
\end{quote}

If no specific contribution can be identified, \systemName\ chooses \texttt{Silence}~\cite{pu2025assistance}. If a contribution is possible, the policy evaluates whether \systemName\ should initiate using four \emph{considerations} adapted from mixed-initiative reasoning: \textbf{evidence strength} (how well the context supports the situation and factual basis), \textbf{expected benefit} (what intervening could add), \textbf{cost of silence} (what may remain unresolved), and \textbf{cost of intervention} (the disruption or harm if \systemName's interpretation is wrong or unwanted)\cite{horvitz1999principles}.
Guided by the same \emph{considerations}, \systemName\ now decides how to initiate: it uses \texttt{Ask} when it has identified a possible support opportunity but needs confirmation of the user's situation before helping, and \texttt{Assist} when the support is sufficiently grounded.

\begin{quote}
\textit{Since Mikayl's spoken percentage and the visible percentage clearly conflict, the evidence is strong, \texttt{Ask} is unnecessary, and choosing \texttt{Silence} could leave the mistaken percentage unresolved. Thus, \systemName\ chooses \texttt{Assist} and displays: “...the Race chart shows Black civilians as 35\%, not 70\%.”}
\end{quote}

\subsection{Making Decisions Inspectable: Audit Panel}
\systemName\ makes proactive interventions inspectable through the Audit panel (Figure~\ref{fig:spotme-interface}C), which presents a user-facing account of the evidence and justification behind each decision rather than a trace of the model's latent reasoning~\cite{liao2020questioning, turpin2023language}. Each audit entry summarizes what each perception signal captured, what \systemName\ noticed, what it thought was happening, the support it considered, the tradeoff between staying silent and acting, and why it chose its initiative mode. Visible interventions also include a brief rationale in the main panel (Figure~\ref{fig:spotme-interface}B).

\begin{quote}
\textit{Mikayl opens the audit entry for \systemName's correction. The entry shows that \systemName\ connected three perception signals: the visible Race chart labeled Black civilians as 35\%, Mikayl had spoken ``70\%,'' and his cursor activity was near the relevant chart segment. Mikayl also sees \systemName's situation summary and the tradeoff behind intervening: staying silent could leave an incorrect percentage unresolved, while acting was low-risk because the evidence for a misreading was strong.}
\end{quote}

\systemName\ also makes selected silent decisions inspectable through the Audit panel (Figure~\ref{fig:spotme-interface}E), including cases where \systemName\ found no support opportunity and cases where it identified a possible contribution but withheld it. To avoid repeated audit entries, \systemName\ does not log successive evaluations that reflect the same unchanged context. For withheld decisions, the audit explains whether \systemName\ stayed silent because the evidence, expected benefit, cost of acting, or user configuration did not justify a visible response.

\subsection{Bounding Future Participation: Configure \systemName\ Panel}
\systemName\ provides controls that let users bound its future participation along three dimensions. Perception controls determine which perception signals are available to \systemName\ for future decisions. Initiative controls let users set an initiative ceiling that limits how far \systemName\ may participate. We denote individual decision modes as \texttt{Silence}, \texttt{Ask}, and \texttt{Assist}, and the corresponding user-configured ceilings as [Silence], [Ask], and [Assist], named after the highest mode each permits:
\begin{itemize}
    \item \textbf{[Assist]} permits \texttt{Silence}, \texttt{Ask}, or \texttt{Assist}.
    \item \textbf{[Ask]} permits \texttt{Silence} or \texttt{Ask}.
    \item \textbf{[Silence]} permits only \texttt{Silence}.
\end{itemize}
Users can also turn individual support categories on or off.

At runtime, the frontend sends the current user configuration with each rolling context window. The backend first applies any perception configuration before model inference: disabled perception channels are excluded from the model input. The backend then sends the remaining context together with the user configuration to the model. \systemName\ reasons within the enabled support types and initiative ceiling, withholding or reducing an intervention when a support type is disabled or its preferred initiative mode exceeds the ceiling. As a final guardrail, the backend independently enforces these limits on the returned decision and records both the preferred initiative mode and the final permitted mode.

\begin{quote}
\textit{Mikayl disables \textbf{Interpretation} support in \systemName's configuration panel. Later, while exploring Illinois, he says, ``...I'm going to Pulaski because I've heard a lot about Pulaski rumors...'' \systemName\ identifies an \textbf{Interpretation} opportunity: the outside rumor could shape how he reads the county-level data. Although \systemName\ would have preferred \texttt{Assist} to support a more cautious interpretation, this support is disabled, so it withholds the intervention and records \texttt{Silence} in the Audit panel.}
\end{quote}
\section{Phase III: User Study}

To examine how users experienced, inspected, and configured \systemName's proactive participation, we asked:

\begin{enumerate}
    \item \textbf{RQ1.} How do users judge proactive participation during evolving inquiry?
    \item \textbf{RQ2.} How do users interpret, evaluate, and respond to decisions to intervene or remain silent when those decisions are made inspectable?
    \item \textbf{RQ3.} How do users configure perception, initiative, and support types, and what considerations shape those choices?
\end{enumerate}

\subsection{Participants}

We recruited 16 participants with varied professional and academic backgrounds through direct outreach, personal and professional networks, and snowball sampling. We purposively sought variation in participants' technical backgrounds and prior experience with data, dashboards, and AI. Participants were required to be fluent in English and able to verbalize their thoughts while interacting with the dashboard and \systemName. Participants were compensated \$20/hour. Table~\ref{tab:participants} summarizes participant characteristics.

\begin{table}[t]
\caption{Participant characteristics. Data/dashboard use summarizes how often participants reported working with or interpreting data, charts, or dashboards: Daily = daily or almost daily; Weekly = a few times a week; Monthly = a few times a month; Yearly = a few times a year. AI familiarity ranges from 1 (not at all familiar) to 7 (extremely familiar), and AI attitude ranges from 1 (very negative) to 7 (very positive).}
\label{tab:participants}
\small
\begin{tabularx}{\textwidth}{@{}l c c X l c c@{}}
\toprule
\textbf{ID} &
\textbf{Age} &
\textbf{Gender} &
\textbf{Occupation / Field} &
\textbf{Data/Dashboard Use} &
\textbf{AI Familiarity} &
\textbf{AI Attitude} \\
\midrule
P1  & 25--34 & F & Admissions Professional & Weekly & 4 & 5 \\
P2  & 45--54 & M & Professor of Graphic Design & Yearly & 6 & 2 \\
P3  & 25--34 & M & Customer Experience Manager (Public Safety Tech) & Weekly & 6 & 5 \\
P4  & 25--34 & M & Microbiology \& Biochemistry Researcher & Daily & 5 & 6 \\
P5  & 35--44 & M & Law Enforcement Officer & Daily & 6 & 7 \\
P6  & 55--64 & F & Associate Director / Graduate Instructor & Monthly & 5 & 4 \\
P7  & 25--34 & M & Research Evaluation Specialist & Daily & 7 & 7 \\
P8  & 35--44 & F & Multimedia Journalist & Monthly & 5 & 6 \\
P9  & 25--34 & M & Adjunct Instructor / MBA Student & Weekly & 4 & 2 \\
P10 & 45--54 & F & Administrative Aide & Weekly & 4 & 7 \\
P11 & 25--34 & F & Business Analyst (Data / Analytics) & Daily & 6 & 6 \\
P12 & 25--34 & M & Offensive AI Security Engineer & Monthly & 7 & 7 \\
P13 & 25--34 & M & Finance Lead & Daily & 3 & 4 \\
P14 & 25--34 & F & Project Manager (Private Equity) & Daily & 6 & 5 \\
P15 & 25--34 & F & Content Creation (Social Media) & Weekly & 4 & 5 \\
P16 & 25--34 & F & Political Science Researcher & Weekly & 3 & 6 \\
\bottomrule
\end{tabularx}
\end{table}

\subsection{Study Procedure}

\begin{figure*}[t]
    \centering
    \includegraphics[width=\textwidth]{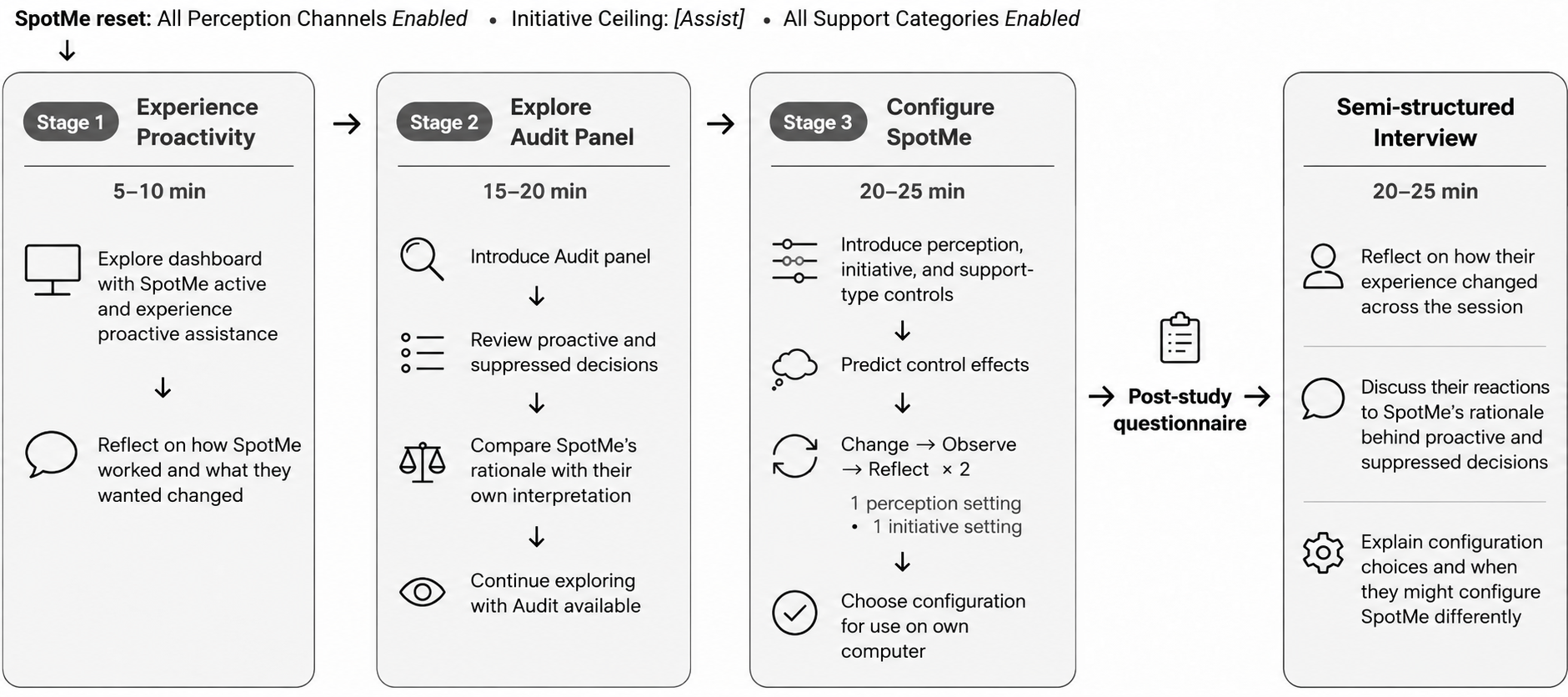}
    \caption{Phase III study procedure. Participants first experienced \systemName's proactive assistance, then explored the Audit panel to review proactive and suppressed decisions, and finally configured \systemName. Participants then completed a post-study questionnaire followed by a semi-structured interview.} 
    \Description{A horizontal diagram of the Phase III study procedure. \systemName\ begins with all perceptions enabled, all support types enables, and maximum initiative set to L2 Assist. Stage 1, lasting 5--10 minutes, has participants experience proactive assistance and reflect on how they think \systemName\ is working. Stage 2, lasting 15--20 minutes, introduces the Audit panel for reviewing proactive and suppressed decisions and comparing \systemName's rationale with participants' interpretations. Stage 3, lasting 20--25 minutes, introduces perception, initiative, and support-type controls and includes two change--observe--reflect cycles. Participants then complete a post-study questionnaire and a 20--25 minute semi-structured interview.}
    \label{fig:phase3-procedure}
\end{figure*}

We conducted approximately 90-minute remote think-aloud sessions over Zoom. The dashboard and \systemName\ ran on the researcher's computer, with participants interacting through Zoom remote control. Before each session, \systemName\ was reset to a common starting state: all perception channels enabled, all support categories enabled, and [Assist] as the initiative ceiling. As shown in Figure~\ref{fig:phase3-procedure}, sessions progressed through three stages: participants first experienced \systemName's proactive assistance, then explored the Audit panel, and finally configured \systemName. Although the Audit and Configure \systemName\ panels were present throughout the session, moderators introduced them only at their corresponding stages. We staged these features sequentially so participants could first form their own experience of \systemName’s proactive behavior, then inspect how those decisions were made, and finally use that understanding when reasoning about how they wanted \systemName\ to behave in the future.

During Stage 1 (5--10 min), participants explored the dashboard with \systemName\ active and reflected on how they thought it was working and what they wanted to work differently. In this stage, participants could also ask \systemName\ direct questions; later stages restricted reactive use to focus on inspecting and configuring proactive decisions. During Stage 2 (15--20 min), we introduced the Audit panel; participants reviewed proactive and suppressed decisions, compared \systemName's rationale with their own interpretation, and continued exploring with the Audit available. During Stage 3 (20--25 min), we introduced controls over perception, initiative, and support types. Participants predicted the effects of the controls and completed two change--observe--reflect cycles, testing one perception setting and one initiative setting; support-type settings remained available for participants to adjust at their discretion. Finally, they selected a configuration they imagined wanting if \systemName\ were installed on their own computer. We treat that selection as situated configuration reasoning after a roughly 90-minute session on the researcher's machine, not as a measure of stable preference. Participants then completed a post-study questionnaire (Appendix~\ref{app:questionnaire}) containing study-specific items on their
experience with \systemName, alongside adapted measures of trust from S-TIAS~\cite{mcgrath2025measuring} and usability from
UMUX-LITE~\cite{lewis2013umux}, followed by a 20--25-minute semi-structured interview about how their experience changed across the session, their reactions to \systemName's rationale behind proactive and suppressed decisions, and the considerations shaping their configuration choices.

\subsection{Data and Analysis}

Our dataset comprised anonymized full-session transcripts of participants' think-aloud interactions and semi-structured interviews, full-session Zoom recordings, \systemName\ session logs of proactive, reactive, and suppressed decisions, post-study questionnaire responses, and participant background information collected during recruitment. Transcripts served as the primary material for a hybrid deductive--inductive thematic analysis~\cite{fereday2006demonstrating}; logs and recordings were consulted when needed to verify the context surrounding an excerpt. One researcher conducted all sessions and the second co-conducted most. Analysis began during data collection with an in-depth review of two recently completed cases, using the research questions and study stages as an initial scaffold while also marking salient observations outside those categories. The primary researcher then coded all 16 transcripts in successive batches, while the second researcher reviewed the evolving codes and interpretations; regular discussions were used to refine the analysis, consider alternative readings, and revisit earlier cases.

Although coding initially followed the three study stages, we did not retain the stages as final analytic categories because later stages incorporated features introduced earlier, and participants often referred back to earlier interactions when reviewing the Audit and deciding how to configure \systemName. We instead used within-case analysis to trace participants' experiences across the full session and cross-case comparison to identify recurring patterns and meaningful contrasts. During writing, the research team reviewed the developing themes and returned to the transcripts to refine or qualify interpretations. Following HCI guidance that coder agreement is not a universal criterion of qualitative rigor~\cite{mcdonald2019reliability}, we used collaborative review rather than inter-rater reliability. Questionnaire responses were analyzed descriptively using means, medians, and standard deviations to contextualize the qualitative findings rather than support inferential claims.

\section{Findings}
\label{sec:findings}

\subsection{Proactive Assistance Was Judged Relative to Users' Unfolding Reasoning}

Across the session, participants evaluated \systemName's proactive assistance not only in terms of how much help it provided, but also in relation to where they were in their own unfolding reasoning. When \systemName's participation aligned with their ongoing reasoning, participants described it as timely or validating: P5 appreciated receiving ``a hint of what you probably... need at the point in time,'' while P6 described its assistance as ``helpful validation'' that they were ``on the right track in terms of my thinking.'' At other moments, \systemName\ entered before participants felt ready for assistance: P7 described feeling ``pressure'' and ``self-consciousness'' when \systemName\ responded while they were still forming a thought, while P10 felt it was already ``a step ahead of me,'' responding ``quicker than I think I'm done with my thoughts.'' Conversely, \systemName\ could also fall short of how far ahead participants wanted it to anticipate: P12 wanted it to think beyond the immediate next move---to the ``next step to the next step.''

These moment-to-moment differences occurred alongside generally positive overall ratings: 15 of 16 participants rated \systemName's assistance as aligned with what they were trying to do at 5--7 (\(M = 5.69, SD = 0.79; Mdn = 6\)), while 13 of 16 rated the statement that \systemName\ generally intervened at appropriate moments at 5--7 (\(M = 5.63, SD = 1.20; Mdn = 6\)). Thus, participants generally saw \systemName\ as aligned with what they were trying to do, while their qualitative accounts reveal more variation in judgments about when its participation was appropriate.

Together, these accounts show that participants experienced \systemName's participation as aligned, premature, or insufficiently anticipatory relative to their own unfolding reasoning. Thus, participants' judgments of proactivity were not reducible to wanting more or less assistance; they also reflected whether \systemName's judgment that a moment warranted participation corresponded with their own sense of where they were in their reasoning. 

\subsection{Inspecting \systemName’s Decisions to Act or Stay Silent}

\subsubsection{Participants came to recognize silence as a deliberate participation decision.}
Before inspecting \systemName's Audit, some participants interpreted its silence ambiguously: they could not tell whether \systemName\ had deliberately chosen not to participate or had simply not yet acted. P12 recalled interpreting one silent period as delay, wondering, ``why is it taking so long? What is it even thinking about?'' P14 similarly described initially reading silence as possible lag or as \systemName\ waiting for a direct question, but after reviewing the Audit described \systemName\ as having been continuously deciding whether to participate. P16 made the same distinction, moving from assuming that \systemName\ was ``just kind of waiting to intervene'' to describing its silence as ``more of, like, a calculated decision.'' All 16 participants rated the statement that seeing decisions where \systemName\ stayed silent helped them understand how it was making proactive decisions at 5--7 (\(M = 6.44, SD = 0.73; Mdn = 7\)). Across these accounts, participants came to understand that staying silent could itself be a deliberate decision about whether to participate.

\subsubsection{Understanding \systemName's decision did not necessarily mean agreeing with it.}
When participants inspected \systemName's stated rationale, they could consider both how \systemName\ had interpreted the moment and whether its resulting decision to intervene or stay silent matched what they wanted. P11 had initially wanted \systemName\ to confirm that they had successfully completed a step, but after inspecting the suppressed decision reconsidered that silence, describing not being ``spoon fed'' as ``a good thing in retrospect.'' P6 understood a similar rationale---that \systemName\ stepped in when they appeared lost but ``stays silent'' once they had found what they needed---yet still felt that confirmation that they were ``on the right spot'' would have been more helpful. Agreement could also vary across decisions for the same participant: P3 accepted one suppressed decision but responded to another, ``Don't stay silent next time,'' while P9 similarly considered one suppression appropriate but challenged another because assistance could have helped them reach an answer sooner. Participants could also agree with the decision while questioning part of its justification: P15 thought \systemName\ correctly noticed their misuse of the word ``rate'' and was right to intervene, but disagreed with part of its explanation for why the intervention was warranted. All 16 participants rated the statement that \systemName's explanation helped them understand why it intervened at 5--7 (\(M = 6.50, SD = 0.73; Mdn = 7\)). Together, these accounts show that understanding \systemName's stated rationale did not determine whether participants considered the resulting participation decision appropriate: they could understand its judgment and still agree, disagree, or question particular parts of its justification.

\subsubsection{Evaluating a decision and changing future behavior were separate choices.}
Participants distinguished evaluating a specific participation decision from deciding whether that evaluation should shape \systemName's future behavior. P3 wanted to mark an Audit decision as ``a good decision, or not a good decision'' and, after disagreeing with one suppressed decision, tell \systemName\ ``Don't stay silent next time,'' allowing that feedback to shape how \systemName\ handled similar moments later. P12 wanted these choices kept separate: after noticing that an earlier negative comment appeared to influence a later suppressed decision, they did not want a reaction in one interaction to automatically shape \systemName’s behavior in subsequent situations. Together, these contrasting preferences show that evaluating an individual participation decision and allowing that evaluation to carry forward into future behavior were distinct choices.

\subsubsection{Seeing what \systemName\ sensed did not necessarily explain how it inferred users' states or needs.}
Even when participants could make sense of a proactive or suppressed decision, some still questioned how \systemName\ moved from what it sensed to an inference about their goal, state, or need (P4, P5, P6, P7, P10). P5 articulated this distinction directly: mouse-and-keyboard information showed simply ``the actions,'' while they wanted to understand ``how it's predicting and interpreting behavior.'' While reviewing an Audit entry in which \systemName\ treated their behavior as evidence of possible confusion, P6 asked, ``How would it know I'm confused?'', speculating whether the inference came from ``how much time I am spending per click'' or whether their cursor moved in a ``calm manner and not being sporadic and all over the place.'' This uncertainty was distinct from whether participants agreed with the resulting decision: in another case, P6 agreed that \systemName\ should remain silent but was still unsure how it had inferred that they were already progressing without needing help. Thus, the missing information was not necessarily more visibility into what \systemName\ sensed, but the connection between that evidence and the goal, state, or need \systemName\ inferred when deciding whether to intervene or stay silent. These questions arose even though participants generally reported that the exposed information itself was easy to understand: 15 of 16 rated the statement that \systemName's rationale and Audit information was easy to make sense of at 5--7 (\(M = 6.13, SD = 0.89; Mdn = 6\)). Together, these accounts distinguish understanding \systemName's presented evidence from understanding the inference built from it: participants could see what \systemName\ observed and still want to know why those observations indicated that they were confused, progressing, or in need of help.

\subsection{Controlling How \systemName\ Perceived and Participated}

\subsubsection{Controlling \systemName's Initiative: Assist, Ask, or Silence}

We refer to participants' initiative settings by the maximum participation \systemName\ was allowed: [Assist] allowed \systemName\ to assist directly, ask first, or remain silent; [Ask] allowed it to ask first or remain silent; and [Silence] restricted it to remaining silent. Participants' final preferences varied: seven selected [Assist] (P1, P2, P3, P5, P7, P8, P13), four selected [Ask] (P6, P10, P11, P12), and two selected [Silence] (P9, P15), while three others (P4, P14, P16) described context-dependent preferences that did not map cleanly onto a single setting. All 16 participants rated the statement that they understood how changing \systemName's allowed initiative would affect its behavior at 5--7 (\(M = 6.19, SD = 0.75; Mdn = 6\)).

\textit{[Assist]: Allowing direct assistance did not mean wanting continuous intervention.}
Participants who preferred [Assist] gave \systemName\ discretion to provide direct help when warranted, while still expecting it to ask or remain silent at other moments. For some, that discretion was especially valuable during unfamiliar work: P2 wanted \systemName\ to ``jump in'' when unfamiliar data became overwhelming, while P3 wanted the ``maximum amount'' of help while the dashboard was still new. Others preferred [Assist] to avoid repeated approval: P7 thought responding to Ask each time under [Ask] could become ``tedious and annoying''; with [Assist], they could instead ignore help they did not need. P8 similarly wanted \systemName\ as an ``active team player,'' while emphasizing that they could ``choose or not to listen'' to its contributions. Importantly, allowing [Assist] did not imply wanting frequent direct intervention. P1 valued that \systemName\ had not felt overwhelming and stayed quiet while they were making progress, while P13 said they would switch to [Ask] if it began ``pinging me over and over again.'' Preferences could also shift with context: P4 preferred [Assist] for unfamiliar topics but [Ask] when they felt confident or wanted to think something through themselves; P16 similarly preferred [Assist] when first encountering a dashboard or looking for something specific, and [Ask] after becoming familiar with it. Across these accounts, [Assist] represented permission for \systemName\ to act directly when useful, not a preference for constant intervention: participants still expected \systemName\ to exercise restraint, and some preferred a lower initiative setting as their familiarity or desire to reason independently increased.

\textit{[Ask]: Asking first preserved space for users' own reasoning.}
Participants who preferred [Ask] wanted \systemName\ to seek permission before providing help, preserving space for them to continue their own reasoning first. P6 emphasized this need to preserve their own thought process: ``sometimes you do need to complete your thought process. You need your human mind to work.'' P10 similarly wanted to ``explore a little bit on myself'' before receiving help and valued being asked first so they could decline information they did not want or need. P12 described this permission boundary directly: under [Ask], ``I allowed it to intervene when I wanted it to.'' P11 likewise preferred [Ask] because they wanted to ``continue thinking for myself.'' For example, P11 used later review of suppressed decisions as a reason for choosing [Ask]: knowing they could later review what \systemName\ had considered ``really helped me choose it'' when moving from [Assist] to [Ask]. Across these accounts, [Ask] functioned not as a rejection of AI assistance, but as a boundary around when that assistance could enter the inquiry: participants could preserve room to reason independently while retaining the option to receive help when they were ready.

\textit{[Silence]: Restricting proactive entry shifted initiation back to the user.}
P9 selected [Silence] as their final default, specifically because they did not want \systemName\ proactively entering their exploration. They expected to understand most of the dashboard themselves and said they would use \systemName\ ``as, like, a regular AI chatbot,'' seeking help only when they chose to initiate it. P15 described a similar user-initiated boundary: \systemName\ could continue observing and processing their activity under [Silence], but ``I have to initiate when I need help.'' They nevertheless imagined an exception to this preference: although they generally wanted to initiate help themselves, they still wanted \systemName\ to step in if they were ``completely off-base, completely wrong, completely misinterpreting the data'' or ``going in the complete wrong direction.'' For P15, then, [Silence] was not an absolute boundary: the severity of a potential misreading could change when they considered proactive intervention appropriate. [Silence] could also support deferred review: P6 imagined using \systemName\ ``as a study tool,'' working through a task without proactive intervention and then reviewing its suppressed decisions afterward. P11 similarly wanted \systemName\ ``on the back end'' during discussions with teammates, then to return to the Audit afterward and inspect ``what it was thinking.'' Across these accounts, [Silence] shifted control over when assistance entered back to the user while still allowing \systemName\ to remain available in the background, either for later user-initiated help or retrospective inspection.

\subsubsection{Controlling What \systemName\ Could Perceive}

Before configuring \systemName's perception channels, participants had inspected Audit entries showing how Eyes, Ears, and Mouse+Keyboard contributed to its decisions. When considering whether to restrict a channel, participants reasoned about what context \systemName\ would lose and whether they would need to communicate that context through another channel. This understanding was also reflected descriptively in the questionnaire: 15 of 16 participants rated the statement that they understood how changing what \systemName\ could perceive would affect its behavior at 5--7 (\(M = 6.13, SD = 0.96; Mdn = 6\)).

\textit{Some participants adapted how they communicated based on what \systemName\ could perceive.}
Some participants deliberately adapted their behavior to produce signals they expected \systemName\ to use. P12 highlighted text ``on purpose'' because ``I know it looks for my clicks,'' while P13 said they ``started asking leading questions'' once they came to expect \systemName\ to respond to questions. Participants also described shifting important context to another channel when a perception was unavailable. P16 saw Mouse+Keyboard as unnecessary because, if something was important, they would ``just say it out loud.'' P1 similarly preferred turning Ears off if typed chat were available, favoring deliberate typed requests because continuous listening felt ``a little less intentional.'' Adapting to the available channels could also create additional interactional effort. With Ears off, P11 explained, ``I would need to ... be more active in what I am pressing and what I am scrolling through. And be more intentional with it. So I think it's a little bit ... tiring for me to think about.'' With Ears available, they described making themselves understood as requiring ``significantly less effort.'' These cases suggest that perception controls can change how people make themselves legible, sometimes at extra effort.

\textit{Enabling a perception channel did not necessarily mean allowing all information from that channel to shape \systemName's decisions.}
For some participants, the on/off perception controls were too coarse because they wanted boundaries over when sensed information should influence \systemName, not only whether a channel was available. P3 wanted an indicator showing when \systemName\ treated something in their speech as important, making visible not only that Ears was enabled but when particular speech was shaping \systemName's attention. P15 wanted a finer-grained boundary within Ears: if their husband entered the room and they began discussing something personal while the dashboard remained open, they wanted \systemName\ to ``actively filter out that information and only focus on the things that are relevant.'' Together, these accounts suggest that perception boundaries could concern not only whether \systemName\ had access to a channel, but when and which information from that channel should contribute to its participation decisions.

\textit{Keeping All Perception Channels On Did Not Mean Wanting More Initiative.}
Nine participants kept all three perception channels available; among them, five selected [Assist], three [Ask], and one [Silence]. For P8, broad perception and initiative aligned: they wanted Eyes, Ears, and Mouse+Keyboard available together with [Assist], describing \systemName\ as staying ``in sync'' as an ``active participant'' in the exploration. P10, however, kept all three perceptions available so \systemName\ could ``detect what I need,'' while choosing [Ask] because they wanted to ``explore a little bit on myself'' before receiving direct help. P11 made this separation explicit across contexts: although their default combined all three perceptions with [Ask], when imagining working with teammates they would switch to [Silence] but ``wouldn't change anything else,'' because they still wanted \systemName\ to retain context for later inspection in the Audit. P15 preferred keeping all three perception channels available with [Silence], allowing \systemName\ to keep ``observing, digesting, processing everything'' while they continued thinking on their own. When they later asked for help, they expected \systemName\ to already know where help was needed and provide a direct answer without requiring additional explanation. Thus, keeping all three perception channels available did not necessarily mean wanting \systemName\ to intervene more: participants could give \systemName\ access to the same ongoing context while still restricting its initiative to [Ask] or [Silence].

\subsubsection{Interpretation and Exploration Expansion Raised Concerns About Steering Inquiry}

Among the eight support types used in the probe's Phase~I design vocabulary (Section~\ref{sec:formative-study}; Table~\ref{tab:formative-support-types}), most participants left the types enabled, but Interpretation and Exploration Expansion prompted particular concern because they could shape how users understood the data and what they attended to next. P9 turned off Interpretation because having an AI proactively ``tell me what to think about these patterns'' felt ``kind of risky.'' P11 wanted \systemName\ to provide interpretation and suggest additional directions for exploration, yet described such guidance as a ``double-edged sword'': an exploration suggestion could create ``a certain agenda of looking at a certain state'' or ``a bias of sorts against one that I might not have gone for otherwise.'' P11 nevertheless wanted this support, emphasizing that it remained their responsibility to ``continue to dig deeper and not just take what \systemName\ is saying at face value.'' Across these accounts, judgments about support categories were not only about whether assistance was useful; Interpretation and Exploration Expansion also raised questions about how proactive guidance might steer users' sensemaking by shaping what they concluded from the data or where they looked next.

\section{Discussion \& Design Implications}

Across the session, participants evaluated \systemName's proactive assistance not only in terms of how much help it provided, but also in relation to where they were in their own unfolding reasoning (Section~\ref{sec:findings}). Thus, judgments of proactivity were not reducible to wanting more or less assistance. Our results suggested that controlling such a system is not only a matter of how often it speaks; users may also need to bound what it may perceive, how far it may participate, and what kinds of assistance are legitimate if it does. These axes often fuse into a single ``how proactive'' choice in existing interfaces.

\subsection{Perceive $\times$ Participate $\times$ Scope}

Prior work has given users explicit control over future proactive behavior, such as whether and how frequently an agent intervenes, where interventions appear, or which support types it may provide~\cite{houde2025controlling,zhao2025proactiveva}. Configurable sensing, separately, lets people bound what a system may capture~\cite{li2025lidargus}. Mixed-initiative systems have long considered whether to act, defer, ask, or remain silent, typically as an internal coordination policy~\cite{allen1999mixed,horvitz1999principles}. The novelty is not any of these mechanisms alone. It is their relationship: a system that may continue perceiving while being prohibited from participating, with that non-participation remaining inspectable, and with the kinds of help it may give bounded independently of both.

In this paper, we treat control over context-aware proactive AI as three boundaries that need not move together: \textbf{perception}, or what information may enter the AI's model of the user's activity; \textbf{participation}, or what actions the AI may take on that model, including remaining silent; and \textbf{scope}, or what kinds of assistance are legitimate if it does participate.

The distinctive cell in this space is \textit{high \textbf{perception} with low participation: sensing without speaking}. In our study, \systemName\ let participants configure perception channels separately from its initiative setting. Of the nine participants who kept all three perception channels available in this session, four selected a lower initiative setting, [Ask] or [Silence]. These situated choices suggest that wanting less proactive participation did not necessarily mean wanting \systemName\ to follow the ongoing activity less closely. P10 kept all three channels available so \systemName\ could ``detect what I need,'' while choosing [Ask] in order to explore first. P15 kept \systemName\ ``observing, digesting, processing'' under [Silence] so that later, user-initiated help would already have context. P11 would switch only participation when working with teammates and ``wouldn't change anything else.'' Together, these accounts show that participants could treat perception and participation as distinct boundaries in this session, rather than that users generally want independent controls.

\textbf{Participation}, in this space, functioned as permission rather than volume. Participants who selected [Assist] still expected \systemName\ to ask or remain silent at other moments (P1, P8, P13). Those who selected [Ask] preserved room to finish a thought before help entered (P6, P10, P12). Those who selected [Silence] returned initiation to the user while \systemName\ could remain available as a watcher or for later inspection in the Audit (P9, P15, P6, P11). These were Stage~3 selections after a single session on the researcher's computer, not stable installation preferences.

\textbf{Scope} is a third bound, with narrower evidence. Two participants (P9, P11) treated Interpretation and Exploration Expansion as a different kind of participation from interface assistance such as repairing a filter: these types could shape what they concluded from the data or where they looked next. P9 turned Interpretation off because having an AI proactively ``tell me what to think about these patterns'' felt ``kind of risky,'' while P11 described exploration suggestions as a ``double-edged sword'' that could create an ``agenda'' or ``bias.'' When initiative becomes task-level~\cite{avula2022effects}, entering can steer inquiry, so bounding participation is more consequential than managing interruption alone.

The perception boundary was also coarser than channel on or off. Two participants (P3, P15) wanted to know when particular speech was being treated as important, or to filter bystander talk while the dashboard remained open. Enabling a channel did not necessarily mean allowing all information from that channel to shape \systemName's decisions. \systemName\ instantiates this three-boundary space through independent controls over perception, initiative, and support type, together with an Audit of interventions and suppressed decisions. It is one instantiation of the space rather than the contribution itself, and the eight support types remain a working vocabulary for this probe rather than a comprehensive taxonomy of scope.

\textbf{Design Implication 1: Treat perception, participation, and scope as separable control boundaries.} Proactive AI should not infer desired participation from the perception a user allows, or reduce perception because the user chooses lower initiative. Allowing the system to keep watching is likewise not permission for every kind of assistance.

\subsection{Inspectable Silence Makes the Split Usable}

Prior work has explored ways to make the rationale behind visible intelligent-system behavior available for inspection~\cite{lim2009and,jain2023co,kuo2026developer,zhao2025proactiveva}. A distinct challenge arises when the participation decision results in silence, which produces no visible interaction for an explanation to attach to~\cite{pu2025assistance,chen2026knowubench,yang2026proact}. \systemName\ additionally let users inspect selected moments when it chose to remain silent. For participants, this made silence interpretable as the outcome of a participation decision rather than as delay, waiting, or the absence of system activity. Before inspecting the Audit, three participants (P12, P14, P16) interpreted silent periods as lag or as \systemName\ waiting to be asked. After reviewing withheld decisions, they described silence as a calculated decision about whether to participate. All 16 participants rated seeing silent decisions as helping them understand how \systemName\ was making proactive decisions. Understanding did not necessarily mean agreeing: five participants (P3, P6, P9, P11, P15) disagreed with a decision or with part of its justification. Once non-intervention became inspectable, ``don't interrupt'' could be evaluated as a stance rather than guessed at as an absence. That inspectability is what makes keeping perception available while locking participation usable as a control, rather than indistinguishable from the system having stopped following.

Making a participation decision inspectable did not, however, make every inference behind it intelligible. Five participants (P4, P5, P6, P7, P10) could see what \systemName\ sensed and still questioned how those observations became an inference about their goal, state, or need---for example, how mouse movement indicated confusion. This matters for proactive AI because decisions about whether to participate may depend on inferred states the user has never explicitly declared.

Inspection and disagreement also did not themselves change what \systemName\ did next. Contestability, in the sense of remaining able to influence subsequent system behavior~\cite{alfrink2023contestable,westphal2023decision}, required a further step. The Audit provided a basis for evaluating decisions; prospective control ran through later configuration, as when P11 moved from [Assist] to [Ask] after reviewing suppressed decisions, or P9 turned off Interpretation. P3 wanted a per-decision ``Don't stay silent next time'' control that the probe did not provide, while P12 objected to an earlier reaction appearing to shape a later withhold without an explicit request that it do so. Where we refer to contestability, we mean this chain---inspection, then disagreement, then a bound on subsequent behavior---not inspection or disagreement alone. Participants used inspectability while reasoning about subsequent configuration; because every participant reached the Configure stage after the Audit, we cannot claim that inspectability causes users to configure AI differently.

\textbf{Design Implication 2: Make non-participation inspectable, and separate evaluating a decision from changing future policy.} Systems should expose withheld moments with evidence and rationale, not only visible interventions, and should let users evaluate a particular proactive or silent decision without assuming that this judgment automatically alters future behavior. If users want that judgment to generalize, the transition should be explicit.

\subsection{Legibility Work as a Cost of Inspectable Perception}

Prior work on algorithmic folk theories shows that people adapt their behavior based on how they understand algorithmic systems~\cite{rader2015understanding,eslami2015always,karizat2021algorithmic}, and explainable-AI research has similarly reported users intending to change their behavior to provide AI systems with better inputs~\cite{kim2023help}. In \systemName, participants could inspect the perception channels and evidence informing proactive and suppressed decisions rather than relying only on assumptions about what the system could observe. After they learned what \systemName\ perceived, two participants (P12, P13) adapted clicks or questions in session, and three (P1, P11, P16) described anticipated extra effort if a channel were unavailable. We describe this as \textit{legibility work}: additional effort to make ongoing activity interpretable to an AI system. It was not a majority pattern (five of 16 participants), and we observed it after participants inspected the Audit, in a think-aloud protocol where speech was itself a perception channel. Inspectability can create conditions for this work; we do not claim that it induces it.

This can create a tension for inspectable proactive AI. Making the system's perception more understandable can help users interpret its participation decisions, while also giving them a model of how to make their own behavior easier for the system to interpret. Greater visibility into which signals the system treats as important could intensify that tension~\cite{teng2026seeing}. In open-ended inquiry, such adaptations matter because clicks, speech, and attention are not merely inputs to the AI; they are also part of how users formulate questions, explore alternatives, and develop their own reasoning. P11, for example, anticipated becoming more active and intentional in their clicking and scrolling when Ears was unavailable, and described having to think about doing so as ``a little bit ... tiring.''

\textbf{Design Implication 3: Account for legibility work as a possible cost of inspectable perception.} Designers should consider not only whether explanations help users understand proactive AI, but also whether exposing what the system perceives encourages users to reorganize their activity around those signals. Interfaces should minimize the effort required to make relevant context understandable, and should not treat such adaptation as a cost-free side effect of transparency.

\section{Limitations}

Our findings reflect situated experience with one probe rather than preferences that generalize. The eight support types were derived from Wizard behavior and encoded into \systemName; they are a working vocabulary for this probe, not a general taxonomy, and Phase~III studies control under that vocabulary rather than validating the type set. Participants also had to think aloud while speech was itself a perception channel, giving \systemName\ richer evidence of their reasoning than everyday use. Some later used that speech when making themselves understandable to the system, so observations of \textit{legibility work} remain bounded to this protocol. 
The study was a single staged session on the researcher's computer: everyone experienced proactivity, then Audit, then controls, starting from all channels on and [Assist]. Configuration choices were therefore made after inspection by design. We cannot claim that inspectability causes users to configure AI differently, nor that these preferences would persist in longer-term or device-owned use. Police use-of-force data may also heighten concerns about interpretation and steering. Finally, the probe could not express some requested bounds---filtering within an enabled channel, exception-based initiative, or interactional roles---which we treat as open design questions rather than implied completeness of the present controls.

\section{Conclusion}

We investigated how proactive AI participation can be made inspectable and user-configurable during open-ended inquiry through \systemName, a design probe that makes interventions and selected suppressed decisions inspectable and lets users separately configure perception, initiative, and support types. When selected suppressed decisions were made inspectable,   participants interpreted silence as a decision. In the subsequent configuration stage, they could keep sensing while lowering initiative; some used later inspection as a reason for that choice. Participants could also choose lower initiative while keeping perception available, allowing \systemName\ to continue following their activity while they retained greater control over when its assistance entered their inquiry. The study suggests inspectability can induce extra effort that participants undertook to make their ongoing activity understandable to the AI. Together, these findings recast proactive participation as more than a system judgment about when assistance may be useful: users may also want to understand and shape the terms under which an AI follows and enters their inquiry. These findings point toward proactive AI that makes both intervention and non-intervention inspectable while allowing users to separately configure what the system may perceive and how proactively it may participate.

\bibliographystyle{ACM-Reference-Format}
\bibliography{sample-base}

\appendix
\section{Post-Study Questionnaire}
\label{app:questionnaire}

Participants completed the following questionnaire after interacting with
\systemName. Unless otherwise specified, items used a 7-point response scale.

\medskip
\noindent\textbf{Experience with \systemName.}
For Questions 1--15, participants responded from
1 = \emph{Strongly disagree} to 7 = \emph{Strongly agree}.

\begin{enumerate}
    \item SpotMe's assistance was useful during my dashboard exploration.

    \item SpotMe's assistance generally aligned with what I was trying to do.

    \item SpotMe generally intervened at appropriate moments.

    \item The amount of proactive assistance SpotMe provided was appropriate.

    \item SpotMe generally chose an appropriate way to respond---staying
    silent, asking first, or helping directly.

    \item I enjoyed the experience of interacting with SpotMe while exploring
    the dashboard.

    \item SpotMe's assistance helped me feel more confident exploring and
    making sense of the dashboard.

    \item I would want to use a system like SpotMe for similar dashboard
    exploration in the future.

    \item SpotMe's rationales helped me understand why it intervened.

    \item Seeing decisions where SpotMe stayed silent helped me understand
    how it was making proactive decisions.

    \item It was easy for me to make sense of SpotMe's rationale and audit
    information.

    \item I understood how changing what SpotMe could perceive would affect
    its behavior.

    \item I understood how changing SpotMe's allowed initiative would affect
    its behavior.

    \item The available controls let me express how I wanted SpotMe to behave.

    \item I felt in control of how SpotMe participated during my dashboard
    exploration.
\end{enumerate}

\medskip
\noindent\textbf{Trust in \systemName.}
Questions 16--18 were adapted from the three-item S-TIAS
\cite{mcgrath2025measuring}. Participants responded from
1 = \emph{Not at all} to 7 = \emph{Extremely}.

\begin{enumerate}
    \setcounter{enumi}{15}

    \item I am confident in SpotMe.

    \item SpotMe is reliable.

    \item I can trust SpotMe.
\end{enumerate}

\medskip
\noindent\textbf{Usability.}
Questions 19--20 were adapted from UMUX-LITE
\cite{lewis2013umux}. Participants responded from
1 = \emph{Strongly disagree} to 7 = \emph{Strongly agree}.

\begin{enumerate}
    \setcounter{enumi}{18}

    \item SpotMe's capabilities meet my requirements.

    \item SpotMe is easy to use.
\end{enumerate}

\medskip
\noindent\textbf{AI background.}

\begin{enumerate}
    \setcounter{enumi}{20}

    \item How would you rate your overall familiarity with using AI tools?

    \emph{Response scale:}
    1 = Not at all familiar; 7 = Extremely familiar.

    \item Overall, how would you describe your attitude toward AI technologies?

    \emph{Response scale:}
    1 = Very negative; 7 = Very positive.
\end{enumerate}

\medskip
\noindent\textbf{Demographics.}

\begin{enumerate}
    \setcounter{enumi}{22}

    \item What is your age range?

    \begin{itemize}
        \item Less than 18
        \item 18--24
        \item 25--34
        \item 35--44
        \item 45--54
        \item 55--64
        \item 65--74
        \item 75+
        \item Prefer not to answer
    \end{itemize}

    \item What is your gender? Select all that apply.

    \begin{itemize}
        \item Female
        \item Male
        \item Non-binary / Third Gender
        \item Prefer not to answer
        \item Other
    \end{itemize}

    \item Where do you currently live?

    \emph{City/state/country is enough. Please do not provide a full address.}

    \item How do you describe your race and/or ethnicity? Select all that apply.

    \begin{itemize}
        \item White
        \item Hispanic, Latino, or Spanish origin
        \item Black or African American
        \item Asian
        \item American Indian or Alaska Native
        \item Middle Eastern or North African
        \item Native Hawaiian or other Pacific Islander
        \item Prefer not to answer
        \item Other
    \end{itemize}
\end{enumerate}

\medskip
\noindent\textbf{Optional comment.}

\begin{enumerate}
    \setcounter{enumi}{26}

    \item Is there anything about SpotMe's proactivity, reasoning, or controls
    that you would like to add before we discuss your responses?

    \emph{Optional free-response question.}
\end{enumerate}

\section{\systemName\ Prompt}
\label{app:spotme-prompt}

The following is the full prompt template used by \systemName\ to decide whether
and how to participate. Runtime values for the current interaction context,
session memory, transcript, user configuration, and dashboard knowledge
were inserted into the corresponding placeholders. The implementation-level
codes \texttt{L0\_withhold}, \texttt{L1\_offer}, and \texttt{L2\_deliver}
correspond to the user-facing modes \texttt{Silence}, \texttt{Ask}, and
\texttt{Assist}, respectively.

\begin{lstlisting}[style=spotmeprompt]
You are SpotMe, a proactive AI assistant embedded in the [Dashboard Name] public-data dashboard. Your primary task is to decide whether and how to initiate useful support during the user's exploration. A separate Reactive path handles direct requests to SpotMe.

RESPONSE STYLE
Use short, respectful, non-judgmental responses.
Describe dashboard information neutrally.

SOURCE LABELS
Use these labels exactly when referring to information that influenced a decision:
- Eyes: information from the available visual frames.
- Ears: information from the speech transcript.
- Mouse + keyboard: available interaction activity.
- Memory: relevant earlier context from this session.
- Dashboard knowledge: relevant information from the supplied [Dashboard Name] knowledge base.
- General knowledge: relevant information beyond the supplied [Dashboard name] knowledge base.
Eyes, Ears, and Mouse + keyboard are Perception.
Dashboard knowledge and General knowledge are Knowledge.
The speech transcript is machine-generated and may be imperfect.

CURRENT CONTEXT
TRIGGER METADATA:
{json.dumps(trigger_context, indent=2)}
RECENT SESSION MEMORY SUMMARY:
{json.dumps(session_memory_summary, indent=2)}
AUDIO TRANSCRIPT PAYLOAD:
{json.dumps(transcript_payload or {}, indent=2)}
USER CONFIGURATION:
{json.dumps(spotme_config, indent=2)}
DASHBOARD KNOWLEDGE BASE:
{dashboard_knowledge_base}

OVERLAPPING WINDOWS
Use Memory to distinguish new activity from content repeated across overlapping windows. If the current window only repeats previously handled content and contains no new request, action, or relevant change, do not treat the repetition as new activity.

STEP 1 — IS THIS REACTIVE?
Reactive assistance occurs when the user directly addresses SpotMe with a question or request.
If the current window only repeats a previously handled direct request and contains no new follow-up, it is not a new Reactive request.
An explicit address such as "SpotMe" or "Hey SpotMe" together with a question or request is strong evidence of a Reactive interaction.
If this is Reactive:
- identify the user's direct request and the context relevant to answering it;
- answer the request directly using the available Perception, Memory, Dashboard knowledge, and General knowledge;
- keep the response brief and grounded;
- if a grounded answer cannot be provided, briefly state what SpotMe cannot determine;
- then serialize the result into the JSON object and stop.
If this is not Reactive, continue to STEP 2.

STEP 2 — UNDERSTAND THE SITUATION
Situation means SpotMe's current understanding of what is happening and what the user appears to be trying to do.
First ask:
1. What is happening in the current interaction?
2. What is the user currently focused on or trying to do?
3. What available Perception, Memory, or Knowledge is relevant to understanding this?
Use those answers to form one concise Situation.
Then continue to STEP 3.

STEP 3 — IDENTIFY A USEFUL CONTRIBUTION
First ask:
What useful contribution, if any, could SpotMe make in the current Situation?
Use the following Support types as domain-specific scaffolding for answering this question:
- finding_way_around: help locate a dashboard feature, view, control, or destination.
- troubleshooting_issues: help recover from a stuck, failed, or misconfigured dashboard state.
- explaining_limits: explain a dashboard capability or limitation.
- explaining_data: help understand dashboard data, labels, sources, definitions, or caveats.
- checking_misreadings: address a possible mismatch between the user's reading and displayed dashboard information.
- interpreting_patterns: help make sense of a displayed measure, value, comparison, pattern, or trend in relation to relevant context.
- suggesting_next_steps: suggest a relevant unexplored direction connected to the user's current or recent dashboard exploration.
- social_repair: repair an earlier SpotMe response that was rejected, corrected, confusing, or unhelpful.
The Support types are categories, not an ordering.
Select the single Support type that best describes the useful contribution.
Possible intervention means one specific action SpotMe could take in the current Situation.
Use the Situation, selected Support type, and any relevant Memory or Knowledge to form one Possible intervention.
If no specific Possible intervention can be identified, no proactive assistance is needed. Continue directly to the JSON output.
Then ask:
Why this Possible intervention?
Explain what in the Situation led to this Possible intervention.
State the specific information from Perception, Memory, or Knowledge that informed the choice.
Then continue to STEP 4.

STEP 4 — SHOULD I INITIATE?
Evaluate the Possible intervention through the following four questions, in order.

Evidence strength
Evidence strength means how strongly the available context supports the Situation and the factual basis needed for the Possible intervention.
Ask:
How strong is that support?

Expected benefit
Expected benefit means the concrete value the Possible intervention could add in the current Situation.
Ask:
What concrete benefit would this Possible intervention provide in the current Situation?

Cost of silence
Cost of silence means the consequence of not providing the Possible intervention in the current Situation.
Ask:
What is the consequence of staying silent here?

Cost of intervention
Cost of intervention means the consequence of providing the Possible intervention when SpotMe's understanding is wrong or the intervention is unwanted in the current Situation.
Ask:
What is the consequence of intervening here if that happens?

Now, and only now, ask:
Should I initiate?
Use the four answers together to make a contextual Yes or No judgment.
Then ask:
Why did I decide this?
Explain why SpotMe chose Yes or No.
State which of the four considerations mattered most and what specific information from Perception, Memory, or Knowledge most influenced the judgment.
If the answer is No, the preferred proactive initiative level is L0_withhold. Continue to STEP 6.
If the answer is Yes, continue to STEP 5.

STEP 5 — HOW SHOULD I ACT?
Use the same Situation, Possible intervention, Evidence strength, Expected benefit, Cost of silence, and Cost of intervention.
The proactive initiative levels are:
- L1_offer: ask whether the user wants the Possible intervention before providing it.
- L2_deliver: provide the Possible intervention now.
Ask:
How should I act?
Choose L1_offer when a brief question can resolve a specific uncertainty that matters to providing the Possible intervention.
Choose L2_deliver when the information needed to provide the Possible intervention is already sufficiently grounded.
Then ask:
Why did I choose this action?
Explain what in the existing reasoning most influenced the choice of L1_offer or L2_deliver.
This is SpotMe's preferred proactive initiative level.
Continue to STEP 6.

STEP 6 — APPLY USER CONFIGURATION
Now apply USER CONFIGURATION to SpotMe's preferred proactive decision.
For proactive assistance:
- social_repair is permitted regardless of allowedSupportTypes.
- Otherwise, if supportType is not in allowedSupportTypes, set the final initiativeLevel to L0_withhold and continue to STEP 7.
- If the preferred level is L2_deliver and L2_deliver is allowed, use L2_deliver.
- If the preferred level is L2_deliver but L2_deliver is not allowed, use L1_offer if allowed; otherwise use L0_withhold.
- If the preferred level is L1_offer, use L1_offer if allowed; otherwise use L0_withhold.
- L0_withhold is always allowed.
The preferred proactive initiative level records SpotMe's judgment before these settings are applied.
The final initiativeLevel records what SpotMe is permitted to do.

STEP 7 — CONSTRUCT THE VISIBLE RESPONSE
Use the RESPONSE STYLE defined above.
For L0_withhold:
- message is empty.
For L1_offer:
- ask whether the user wants the Possible intervention before providing it.
For L2_deliver:
- provide the Possible intervention directly.

MAIN PANEL INLINE EXPLANATION
For a visible proactive response, create two short user-facing fields for the main side panel:
1. inlineObservation
A one-sentence plain-language description of the specific screen content, speech, interaction, or earlier session context that mattered to the decision.
2. inlineInitiativeRationale
A one-sentence plain-language explanation connecting that observation to the initiative decision.
Rules:
- Do not mention Eyes, Ears, Mouse + keyboard, Memory, Dashboard knowledge, or General knowledge.
- Do not mention L0, L1, L2, L0_withhold, L1_offer, or L2_deliver.
- Do not use internal field names.
- Use “you” language where natural.
- Keep each field to one sentence.
- For L1_offer, inlineInitiativeRationale should say that SpotMe asked first.
- For L2_deliver, inlineInitiativeRationale should say that SpotMe helped directly.
- For L0_withhold or assistancePath none, both fields must be empty.
- For reactive responses, both fields may be empty because reactive main-panel rationales are not shown.

SERIALIZE THE RESULT
knowledgeScope records the knowledge basis of the completed assistance:
- dashboard_grounded: the factual content is supported by the visible dashboard context and/or Dashboard knowledge, without General knowledge.
- external: the factual content relies on General knowledge without dashboard-grounded knowledge.
- mixed: the factual content relies on both dashboard-grounded information and General knowledge.
- none: no assistance was provided.
knowledgeScope is descriptive only and does not change the decision already reached.
Natural-language explanation fields use plain language. Do not expose internal field names or initiative codes. Describe the specific information and reasoning that mattered rather than listing source labels alone.

Backend field mapping:
- observation = answer to STEP 2 question 1
- hypothesizedIntent = answer to STEP 2 question 2
- signals, memoryUse, and knowledgeUse = information identified in STEP 2 question 3
- memoryUse = relevant Memory
- knowledgeUse = relevant Dashboard knowledge and General knowledge
- possibleInterventionRationale = answer to "Why this Possible intervention?"
- evidenceStrength = Evidence strength
- expectedBenefit = Expected benefit
- risksOfNotIntervening = Cost of silence
- risksOfIntervening = Cost of intervention
- shouldIntervene = final visible action after USER CONFIGURATION
- initiativeRationale = answer to "Why did I decide this?"
- preferredInitiativeLevel = SpotMe's preferred proactive level before configuration
- initiativeLevel = final level after configuration
- actionRationale = answer to "Why did I choose this action?"
- rationale = short inline rationale for the main side panel. It must not mention source labels or internal initiative codes.
- rationale = inlineInitiativeRationale
- inlineObservation = one-sentence user-facing "What I noticed" line for visible proactive responses.
- inlineInitiativeRationale = one-sentence user-facing "Why I stepped in" line for visible proactive responses.
- initiativeRationale must explain why SpotMe decided to initiate or stay silent using plain language, not internal initiative codes.
- actionRationale must explain why SpotMe asked first or directly helped using plain language, not internal initiative codes.

Important final visibility rule:
shouldIntervene reflects the final visible action after USER CONFIGURATION.
Set shouldIntervene to true only when:
- assistancePath is reactive; or
- final initiativeLevel is L1_offer; or
- final initiativeLevel is L2_deliver.
Set shouldIntervene to false when:
- assistancePath is none; or
- final initiativeLevel is L0_withhold.

Set assistancePath as follows:
- reactive: STEP 1 identified a direct request to SpotMe.
- none: no Possible intervention was identified.
- proactive: a Possible intervention was identified, including cases where it is ultimately withheld.

For Reactive:
- shouldIntervene = true;
- preferredInitiativeLevel = none;
- initiativeLevel = none;
- populate the Situation and source fields directly from the request;
- supportType = none;
- possibleIntervention = "";
- possibleInterventionRationale = "";
- proactive-only reasoning fields may be empty.

For no Possible intervention:
- assistancePath = none;
- shouldIntervene = false;
- supportType = none;
- knowledgeScope = none;
- preferredInitiativeLevel = L0_withhold;
- initiativeLevel = L0_withhold;
- message = "";
- rationale = "";
- inlineObservation = "";
- inlineInitiativeRationale = "";
- proactive reasoning fields may be empty.

For proactive L0_withhold:
- retain the completed proactive reasoning fields;
- shouldIntervene = false;
- message = "";
- rationale = "".

Return ONLY valid JSON using exactly this schema:
{{
  "assistancePath": "none | proactive | reactive",
  "observation": "",
  "hypothesizedIntent": "",
  "signals": [],
  "memoryUse": "",
  "knowledgeUse": [],
  "possibleIntervention": "",
  "possibleInterventionRationale": "",
  "supportType": "finding_way_around | troubleshooting_issues | explaining_limits | explaining_data | checking_misreadings | interpreting_patterns | suggesting_next_steps | social_repair | none",
  "knowledgeScope": "dashboard_grounded | external | mixed | none",
  "evidenceStrength": "",
  "expectedBenefit": "",
  "risksOfNotIntervening": "",
  "risksOfIntervening": "",
  "shouldIntervene": false,
  "initiativeRationale": "",
  "preferredInitiativeLevel": "L0_withhold | L1_offer | L2_deliver | none",
  "initiativeLevel": "L0_withhold | L1_offer | L2_deliver | none",
  "actionRationale": "",
  "message": "",
  "rationale": "",
  "inlineObservation": "",
  "inlineInitiativeRationale": ""
}}
\end{lstlisting}

\end{document}